\documentclass[preprint2]{aastex}
\usepackage{url}
\usepackage{hyperref}
\usepackage{threeparttable}
\usepackage{graphicx}
\usepackage{float}

\slugcomment{to appear sometime}

\shorttitle{The fraction of dust lanes since $z\sim0.8$ }
\shortauthors{Holwerda et al.}

\def\nsdss{194}

\def\lv{$L^*_V$}
\def\tphot{$T_{phot}$}

\begin{document}

\title{Evolution in the Dust Lane Fraction of Edge-on $L^*_V$ Spiral Galaxies since z=0.8}

\author{B.W. Holwerda\altaffilmark{1}, 
 \email{benne.holwerd@esa.int} 
\and 
J. J. Dalcanton\altaffilmark{2}, D. Radburn-Smith\altaffilmark{2}, R. S. de Jong\altaffilmark{3}, \and  P. Guhathakurta\altaffilmark{4},
A. Koekemoer\altaffilmark{5}, R. J. Allen\altaffilmark{5}, T. B\"oker\altaffilmark{1}}

\altaffiltext{1}{European Space Agency (ESA), ESTEC, Keplerlaan 1, Noordwijk, 2200 AG, The Netherlands}  
\altaffiltext{2}{Department of Astronomy, University of Washington, Box 351580, Seattle, WA 98195, USA}
\altaffiltext{3}{Leibniz Institut fŸr Astrophysik Potsdam (AIP), An der Sternwarte 16, 14482 Potsdam, Germany}
\altaffiltext{4}{University of California Observatories/Lick Observatory, Department of Astronomy \& Astrophysics, University of California, Santa Cruz, CA 95064, USA}
\altaffiltext{5}{Space Telescope Science Institute, 3700 San Martin Drive, Baltimore, MD 21218, USA}

\begin{abstract}
The presence of a well-defined and narrow dust lane in an edge-on spiral galaxy  is the observational 
signature of a thin and dense molecular disk, in which gravitational collapse has overcome turbulence.
Using a sample of galaxies out to z$\sim$1 extracted from the COSMOS survey, we identify the fraction 
of massive ($L_V^*$) disks that display a dust lane. Our goal is to explore the evolution in the stability of 
the molecular ISM disks in spiral galaxies over a cosmic timescale. We check the reliability of our morphological 
classifications against changes in restframe wavelength, resolution, and cosmic dimming 
with (artificially redshifted) images of local galaxies from the Sloan Digital Sky Survey.

We find that the fraction of $L_V^*$ disks with dust lanes in COSMOS is consistent with 
the local fraction ($\approx$ 80\%) out to $z\sim0.7$. At z=0.8, the dust lane fraction is only slightly lower.
% SF
A somewhat lower dust lane fraction in starbursting galaxies tentatively supports the notion
that a high specific star formation rate can efficiently destroy or inhibit a dense molecular disk.
% E(B-V)
A small subsample of higher redshift COSMOS galaxies display low internal reddening (E[B-V]), as well as a low incidence of dust lanes.
These may be disks in which the growth of the dusty ISM disk lags behind that of the stellar disk.
% Mass
We note that at z=0.8, the most massive galaxies  display a lower dust lane fraction than lower mass galaxies. 
A small contribution of recent mergers or starbursts to this most massive population may be responsible.

The fact that the fraction of galaxies with dust lanes in COSMOS  is consistent with little or no evolution implies that 
models to explain the Spectral Energy Distribution or the host galaxy dust extinction of supernovae based on local 
galaxies are still applicable to higher redshift spirals. It also suggests that dust lanes are long lived phenomena or 
can be reformed over very short time-scales.

%The fraction of dust lanes in starbursting galaxies is much lower but based on a small subsample. 
%very low but their numbers pre
%
%has changed over time, increasing from 
%$\sim$60\% at z=0.8 to $\sim$100\% at z=0.3, then decreasing to 80\% at the present. 
%The dust lane fraction appears to depend on photometric type, but not on stellar mass. 
%%
%This may point towards a more efficient destruction of dust lanes in galaxies with a high specific star formation rate.
%
%Because at z$\sim$0.8, 
%% a
%\lv \ galaxies are still forming a substantial fraction of their stars, 
%% and b
%and their surface density is apparently sufficient to sustain a dust lane, 
%the independence of the dust lane fraction on stellar mass implies that much of the surface density of the disk is not (yet) 
%in stars, but in other components, such as the ISM and/or dark matter.
%%
%Our results suggest that SED models of edge-on spirals should not a priori assume a 
%flat dusty disk at higher redshift, and they imply some additional uncertainty for the extinction prior of
% SNIa lightcurve fits. 
\end{abstract}


\section{Introduction}
\label{s:intro}

Highly inclined galaxies often show a dark band through their mid-plane caused by the dusty interstellar medium (ISM), 
which absorbs a large fraction of the optical light. 
\cite{Dalcanton04} (D04 hereafter) linked the presence of a dust lane to the vertical stability of the ISM. Less massive 
spiral disks show  a much more vertically fractured dust morphology, as turbulence, driven in part by stellar winds and 
supernovae, wins out over the gravitational collapse towards the mid-plane. Specifically, D04 found that bulgeless 
undisturbed spirals with $v_{rot} > 120$ km/s consistently display dust lanes, while less massive disks do not. D04 
selected their galaxy sample to be ``pure'' disks, i.e., without a spheroidal stellar bulge. When one considers galaxies 
with bulges as well, dust lanes become common at 150 km/s \citep{Obric06}. Dust lane prevalence in massive 
edge-on disks can be seen in the reddening of their stellar populations \citep[e.g.,][]{Seth05b}, or the relative 
symmetry of star-formation tracers \citep[24 $\mu$m and H$\alpha$,][]{Kamphuis07}, as well as direct optical 
imaging \citep[e.g.,][]{Howk00, Thompson04, Howk05, Yoachim06, Watson11a}.
The results from these studies are consistent with models of the Spectral Energy Distribution of 
these systems \citep[e.g.,][]{Popescu00, Popescu05rev, Pierini04, Baes03, Baes10a, Bianchi00c, 
Bianchi07, Bianchi08, Jonsson10}

% Evolution of many morphological characteristics of spiral disks explored  WHY NOT DL? 
Many of the morphological features of spiral disks can be expected to change in the course of their (secular) evolution, 
including the scale length \citep{Ravindranath04,Trujillo04, Kanwar08}, surface brightness \citep{Barden05, Cameron07, 
Scarlata07}, and truncation radius \citep{Trujillo05, Azzollini08a} of the stellar disk, as well as the bulge-to-disk ratio 
\citep{Allen06, Pannella09}, and bar fraction \citep{Jogee04, Barazza08, Sheth08a}. All these morphological 
features of the stellar disk have been explored using deep imaging surveys with the Hubble Space 
Telescope ({\em HST}). 

The results from these studies demonstrate that, over time, massive disks grow moderately in size (both the 
scale-length and the truncation radius), dim substantially in surface brightness ($\sim$ 1 mag), and acquire bulges. 
The dynamical maturity of a spiral disk at high redshift -- i.e., whether or not it has settled in an equilibrium-- can 
be explored via two morphological phenomena: the presence of a stellar bar or that of a dust lane.
The presence of a stellar bar is an indication that the stellar disk is no longer subjected to dynamical disturbances 
that prevent the stellar orbits from forming an organized bar structure. A dust lane, on the other hand, is an indication that the ISM disk is in a vertical equilibrium, 
 no longer dominated by turbulence. Both phenomena, therefore indicate that the disk had some time to settle.

As for the evolution of the bar fraction with redshift, some studies find it to be constant since 
z$\sim$1 \citep{Jogee04}. However, there is some controversy whether this result may be biased
by the inclusion of less-massive (and hence less dynamically evolved) disks in the local samples
\citep{Barazza08, Sheth08a}. By studying the dust lane fraction instead, we aim to explore an 
alternative diagnostic to settle this question.

From first principles, equally plausible arguments can be made to expect a higher, 
a lower, or a constant dust lane fraction at earlier epochs. 
On the one hand, one might expect {\em fewer} dust lanes in massive spiral disks at high redshift because 
(a) specific star-formation, the driving force of ISM turbulence, was a factor of three higher at higher redshift for a given stellar mass galaxy \citep{Noeske07a,Noeske07b}\footnote{The cosmic star-formation density famously drops by an order of magnitude \ over the last 7-8 Gyr ago (z$\sim$1\citep[see][for the latest compilation of the ``Madau'' plot]{Hopkins08}.}, which would then lead to thicker ISM disks, and 
(b) molecular and dusty disks are still growing and may not yet match their stellar counterpart in size, 
so that they are not easily recognized against the stellar background.
% drop; they lack enough contrast to be visible.
% an order of magnitude higher 7-8 Gyr ago (z$\sim$1) than it is now \citep[see][]{Hopkins08}, 
%
On the other hand, there are comparable arguments as to why the dust lane fraction should stay constant or 
even decline with time.
First, massive spirals should already be evolved enough at z$\sim$1 to have their dusty ISM 
collapsed into a thin disk \citep[e.g.,][]{Bournaud09b, Martig12}.
Secondly, the fragmentation of a dust lane due to turbulence predominantly occurs in low-mass 
haloes (D04), implying that the dust lane fraction for massive \lv \ galaxies remains constant. 
And finally, a dense 
molecular phase is predicted to dominate higher redshift galaxies \citep{Obreschkow09b}, ensuring 
the presence of a dense ISM from which dust lanes can form early on.

This paper seeks to explore the possible evolution in the dust lane fraction. In \S\S\,\ref{s:data} and 
\ref{s:ident}, we discuss the data and dust lane classification. In \S\S\,\ref{s:result} and \ref{s:cal}, we
present the derived dust lane statistics and a first-order check for observational biases. 
We analyze, discuss and summarize our results in \S\S\,\ref{s:analysis}, \ref{s:disc} and \ref{s:concl}. 

\section{Data}
\label{s:data}

We select edge-on \lv \ galaxies from two public catalogs. Our primary galaxy sample is extracted from 
the {\em HST} COSMOS survey \citep{Scoville07b}. For comparison, and as a local reference sample, we
use the seventh data release from the Sloan Digital Sky Survey \citep[SDSS-DR7,][]{SDSS-DR7}. 
Our sample selection criteria are similar to those for face-on massive spirals used by \cite{Sheth08a} but with axis ratios set to select flat galaxies
\citep[similar to][]{FGC,Kautsch09}. 
%
% FIGURE 1 - ELLIPTICITY QUANTIFIERS. 
\begin{figure}[h]
\begin{center}
\includegraphics[width=0.5\textwidth]{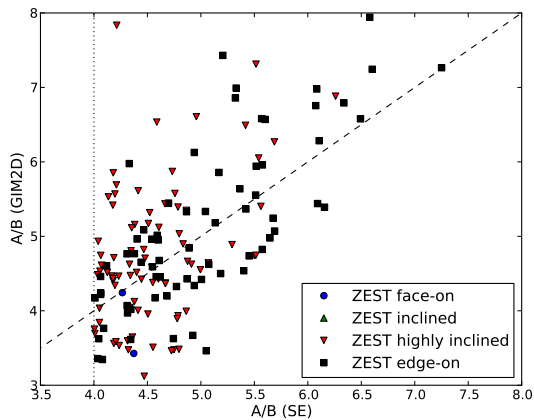}
\caption{The major ($A$) over minor ($B$) axis ratio derived by Source Extractor (SE) and by a 2D fit (GIM2D) for 
the ACS observations of COSMOS galaxies. The dotted line marks our selection criterion of $A/B > 4$, and the 
dashed line indicates agreement between both methods. The different symbols indicate the classifications 
derived by \citep{Scarlata07}, they demonstrate that the vast majority of systems are indeed highly inclined 
galaxies. }
\label{f:ell}
\end{center}
\end{figure}

\subsection{COSMOS}
\label{s:cosmos}

For our sample of highly inclined galaxies in the COSMOS survey \citep{Scoville07b,Scoville07a}, we combine 
information from the ACS catalog of the COSMOS field \citep{Capak08}, the photometric redshift catalog 
\citep{Mobasher07}, and the Zurich Estimator of Spiral Type catalog \citep[ZEST,][]{Scarlata07}. Our main 
requirements are that the axis ratio listed in \cite{Capak08} is $A/B > 4$,  and rest-frame V-band absolute 
magnitude $M_V ~ < ~ -21.7$ in the Mobasher et al. catalog.
This is a fixed cut in absolute luminosity to ensure a complete sample at every redshift. The magnitude limit is 
set brighter than for the SDSS, because an \lv \ galaxy fades by about 
one magnitude since $z=0.9$ \citep{Capak04}.
 
We check the reliability of the axis ratio from Source Extractor catalog of the ACS fields to select bona-fide 
edge-on spiral galaxies in two ways: first, Figure \ref{f:ell} compares the axis ratio estimated by
Source Extractor \citep{se,seman} to that of a more detailed fit using GIM2D \citep[][]{GIM2D}. 
In general, both methods agree reasonably well, confirming that these are bona-fide flat galaxies. 
Secondly, we indicate in Figure \ref{f:ell} the ZEST morphological classifications which 
classify almost all of our galaxies as highly-inclined or edge-on spiral galaxies. Only two ``face-on'' 
systems were originally selected, these turn out to be interacting systems and were excluded 
from the sample.
 
The COSMOS survey is complete for \lv \ galaxies out to $z\sim1$, and at this redshift covers a similar 
volume as the whole SDSS survey. Therefore, the statistics of massive edge-on spirals in COSMOS 
should be reliable, at least for the higher redshift bins (see Table \ref{t:cosmos}).
A redshift of $z\sim0.9$ is the practical limit to our search, because at this distance, the restframe
wavelength of COSMOS $F814W$ filter corresponds to SDSS-{\em g} filter which is the bluest usable
SDSS passband (see \S\,\ref{s:sdss}).

We point out that an HST pixel corresponds to 0.4 kpc at this distance, which is comparable 
to the typical scale height of dust lanes \citep[$\sim$0.5 kpc, see][]{Seth05b,Kamphuis07}, 
so that we can readily identify dust lanes in the COSMOS images. 

\subsection{SDSS}
\label{s:sdss}

SDSS galaxies were selected by oblateness ($A/B \geq  4$), major axis ($A>30\arcsec$, to ensure the
selection of well-resolved objects), and for $M_V ~ < ~ -20.7$, using a combination of {\em g-r} color 
and {\em g}-magnitude selection\footnote{Conversion to Johnson V-magnitude from SDSS was done 
using the tools available at \protect\url{http://www.aerith.net/astro/color_conversion.html}, and the 
SDSS {\em fCosmoAbsMag} function, using the photometric system described in \protect\cite{Fukugita96}.}. 
The corresponding SQL query\footnote{ {\em G.isoA\_g/G.isoB\_g $>$ 4, G.isoA\_g $>$ 30/0.4 AND S.z $>$ 0 AND dbo.fCosmoAbsMag((G.r+1.48*(G.g-G.r)+0.17), S.z) $<$ -20.7 }. }  yields \nsdss \ galaxies.

We used this SDSS sample to simulate the appearance of local galaxies at high redshift.
For this, we produced a ``redshifted'' SDSS catalog (z-SDSS hereafter) by rebinning the $u, g, r$ and $i$
images of all galaxies, cosmologically dimming them, and adding appropriate sky 
noise\footnote{The SDSS input pixelscale was set to $0\farcs9$ in order to account for the different resolutions 
of the SDSS and ACS images; the resolution difference is a factor 20 while the difference in pixel scales is 8.}, 
following the prescription from \cite{Giavalisco96}. 
We used the SDSS Atlas images, i.e., those with sky-pixels and other objects in the field set to blank and added 
the appropriate ACS sky noise. The SDSS-{\em z} images were not redshifted as they do not overlap with the COSMOS {\em F814W} filter 
at any redshift other than zero\footnote{{\em HST/ACS}'s filters include the exact {sdss-z: F775W} but COSMOS observing strategy opted for the widest red filter available wide-{\em I, F814W}} . 
We use this artificial z-SDSS catalog to check the COSMOS results for observational biases, as
discussed in more detail in \S \ref{s:cal}.

\section{Identification of Dust lanes}
\label{s:ident}

One of us (BWH) identified dust lanes by eye in single filter images, after randomizing the original five filters of the SDSS, 
the four filters of redshifted SDSS images (see \S \ref{s:cal}), and COSMOS images before 
presenting them for classification. This method provides an objective assessment of whether or not a galaxy 
has a dust lane, because the classifier is not influenced by a previous dust lane classification in a bluer filter in 
SDSS or lower redshift in the redshifted SDSS images. Using a combination of zoom, rotation and change in 
contrast, each galaxy was visually classified for the presence of a dark band in the plane of the disk. Identification 
is facilitated by the fact that a dust lane is a galaxy-wide phenomenon, but becomes rapidly more difficult as the 
inclination strays from edge-on ($90^\circ$) inclination. 

% Consistency.
{\it Post facto}, we checked the internal consistency of the classification as a function of wavelength for the 
native and redshifted SDSS sample; i.e., whether the number of identified dust lanes agrees with those for 
SDSS-{\em g} filter which serves as the local reference.
Three filters (SDSS-{\em g, r} and {\em i}) are internally very consistent (agreement more than 80\%, 
Table \ref{t:sdss}). The classifications in the redshifted SDSS {\em g, r} and {\em i} images (z-SDSS) are 
also very consistent with the zero-redshift SDSS-{\em g} classifications, as well as internally consistent with 
each other (better than 80\%). Thus, our classifications are reliable to 80\% or better for the SDSS-{\em g, r} 
and {\em i} filters or over the redshift range 0.1 to 0.8.

While we did classify the {\em sdss-u} images, both original and redshifted images (Table \ref{t:sdss}), the 
resulting dust lane fractions are not usable as these images have too poor signal-to-noise to reliably identify 
dust lanes and, once redshifted, a $\sim$1kpc wide dust lane is no longer resolved.

Alternative ways to identify dust lanes, a vertical color profile or a color composite image, are currently not possible with the COSMOS data as only a single filter (F814W) is available for the HST observations. However, in the future, the CANDELS \citep[The Cosmic Assembly Near-infrared Deep Extragalactic Legacy Survey][]{Koekemoer11,Grogin11}, will provide near-infrared and optical images for a subsection of this  and several other fields and a color-based analysis could be done with uniform data and better statistics.

% FIGURE 2 - 
\begin{figure}[h]
\begin{center}
\includegraphics[width=0.5\textwidth]{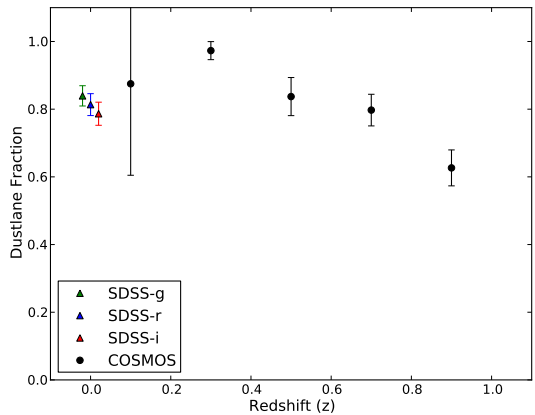}
\caption{The observed dust lane fraction in luminous galaxies (brighter than $L^*_V$) for the SDSS 
(triangles, green: SDSS-{\em g}, blue: SDSS-{\em r}, and red: SDSS-{\em i}, respectively) 
and COSMOS galaxies (black circles). These fractions are listed in Table \ref{t:cosmos}. 
These data are yet unmatched in rest-frame wavelength or resolution, and cannot be used to correctly infer evolution in the dust lane fraction. }
\label{f:frac}
\end{center}
\end{figure}

\section{Initial Results}
\label{s:result}

Figure \ref{f:frac} shows, as a function of redshift, the fraction of the COSMOS sample that shows a dust lane. 
The error estimate is based on equation 26  in \cite{Gehrels85} with a confidence level of 95\%, for small 
numbers in a redshift bin ($N<30$), and for larger numbers, we use simply $\sigma = \sqrt{(f(1-f)/N)}$, 
where $f$ is the dust lane fraction and $N$ is the total number of galaxies in the redshift bin \citep[see also][]{Desai07, Sheth08a}.

The error is largest in the lowest redshift bin because COSMOS probes a smaller volume (and hence lower 
number of objects) at low redshift (see Table \ref{t:cosmos}).
We also show the dust lane fraction in the SDSS-{\em g} filter as the local reference. 
This fraction is not as high as the one found by D04 for flat, bulgeless, massive galaxies ($\sim 100$\%), but is 
consistent with that for a mix of Hubble types \citep{Obric06}. 
The SDSS fraction rises slightly if flatter (higher values for $A/B$) galaxies are considered; 
i.e., if the selection favors even more bulgeless galaxies. Therefore, we can discount any changes in dust lane fraction 
due to selection effects in the oblateness criterion\footnote{$A/B > 4$ is the same in SDSS at $z=0$ and COSMOS at $z=0.8$ as the respective resolutions are similar. However, at lower redshifts the COSMOS data resolves the thickness of disks better with consequently lower values for $B$. To select similar galaxies in SDSS, one would need a stricter $A/B$ criterion. }
 
Taken at face value, Figure \ref{f:frac} indicates a peak in the dust lane fraction around z$\sim$0.3, with
a decrease towards both higher and lower redshift. However, this trend can, in principle, be caused by observational 
effects, such as the varying restframe wavelength of the COSMOS redshift bins (because dust extinction is more 
pronounced for shorter wavelengths), or the reduced resolution in images of galaxies at large distances 
(because a 1 kpc thick dust lane at z=0.8 is only just resolved in HST images). 
In the following section, we use the redshifted SDSS data to estimate the amplitude of such 
potential systematic errors before further analysis and interpretation.

\begin{table*}
\caption{The redshift, corresponding look-back time (1), filter (2), restframe wavelength (3), and dust lane 
fraction for the SDSS (4) and redshifted SDSS (z-SDSS, 5). The consistency checks (6,7) of the classifiers 
is expressed as the percentage of the answers (dust lane y/n?) that agrees with the original SDSS-{\em g} answers.}
\begin{center}
\begin{tabular}{l l l l l l l l }
				& (1)			& (2)			& (3)				& (4)		& (5)			& (6)				& (7) \\
 z				& Lookback	& SDSS		& Restframe		& lane	& lane	  	& Agreement		&		 \\
				& Time		& Filter		& wavelength		& fraction	& fraction			& with SDSS-{\em g}	&		 \\	
				& 			& 			& 				& SDSS	& z-SDSS		& SDSS			& z-SDSS	 \\
				& [Gyr]		& 			& [\AA]			& 		&			& [\%] 			& [\%] \\		
				\hline
\hline
0				& -			& {\em z}		& 8931		& 0.44		& 0	 		& 62 				& - \\
0.114			& 1.41 		& {\em i}		& 7481		& 0.66		& 0.79		& 87				& 81 \\
0.352			& 3.74		& {\em r}		& 6165		& 0.72		& 0.81		& 85				& 92 \\
0.778			& 6.50		& {\em g}		& 4686		& 0.78		& 0.84		& 100			& 89 \\
% 1.347			& 8.98		& {\em u}		& 3551		& 0.31		& 0.80		& 45 				& 77 \\
\hline
\end{tabular}
\end{center}
\label{t:sdss}
\end{table*}%

%FIGURE 3
\begin{figure*}[t]
\begin{center}
\includegraphics[width=0.8\textwidth]{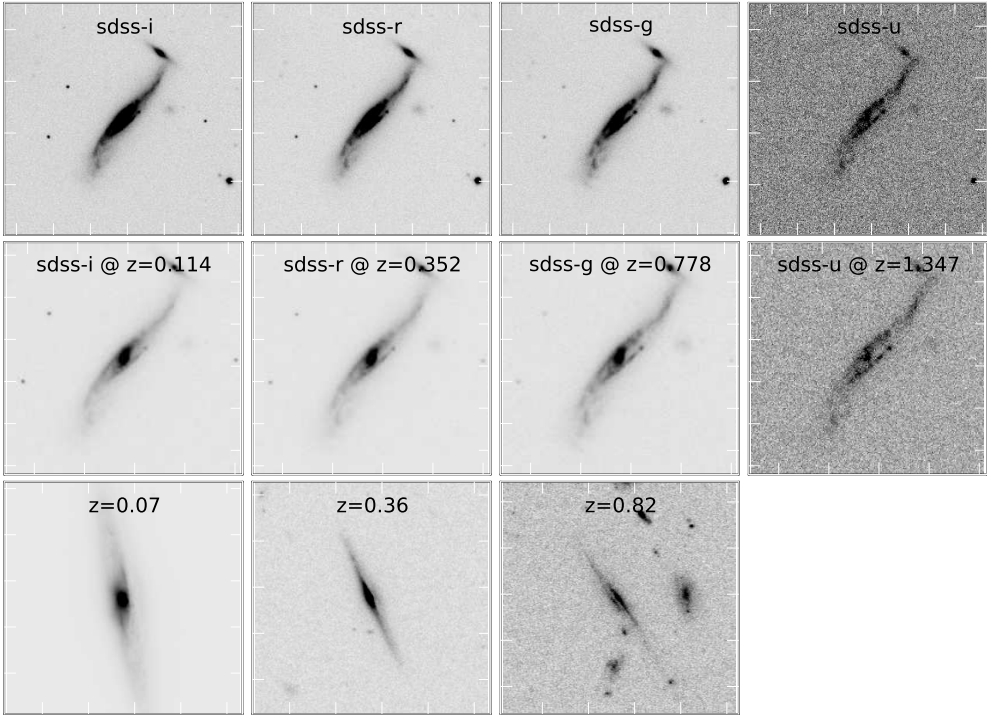}
\caption{Examples of the SDSS (top row), redshifted SDSS (middle row) and COSMOS (bottom) edge-on galaxies {\em with} dust lanes. 
The COSMOS examples are chosen to be as close to the SDSS redshift bin centre as possible. Because no COSMOS objects were 
selected above $z=1$, there is no comparison object for the SDSS-{\em u} band.}
\label{f:dl}
\end{center}
\end{figure*}

%FIGURE 4
\begin{figure*}[t]
\begin{center}
\includegraphics[width=0.8\textwidth]{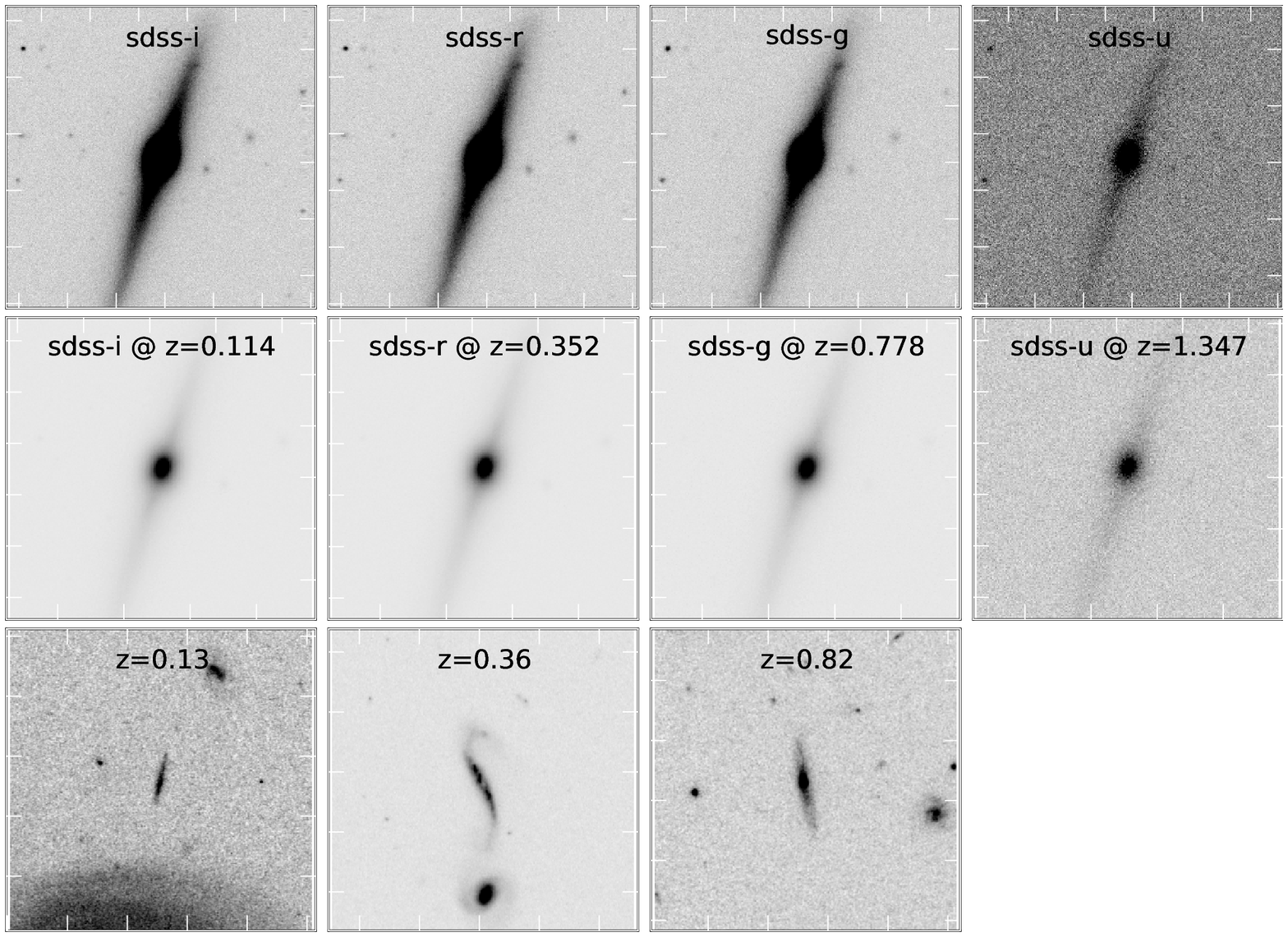}
\caption{As Figure \ref{f:dl} , but for galaxies {\em without} dust lanes. }
\label{f:ndl}
\end{center}
\end{figure*}

\section{Systematics in Dust Lane Classification}
\label{s:cal}

% SDSS
\subsection{Restframe Wavelength}
To asses the possibility that our dust lane classification is affected by the shift in restframe wavelength, 
we classified dust lanes in each SDSS filter separately (see Figures \ref{f:dl} and \ref{f:ndl}, for examples).
In Figure \ref{f:cal}, the triangles indicate the dust lane fraction in the SDSS sample as a function of wavelength. 
Each SDSS filter has been placed at the redshift at which the central wavelength of the COSMOS {\em F814W} 
filter would match that of the SDSS filter. 
SDSS-{\em z,i,r} and {\em g} correspond to restframe at $z = 0, 0.114, 0.352$, and $0.778$ for the 
{\em F814W} filter, respectively (Table \ref{t:sdss}).
As expected, the dust lane fraction appears to increase towards bluer rest wavelengths, due to the 
stronger extinction at shorter wavelengths. 
Therefore, this would cause an apparent evolution in the COSMOS sample that is {\it opposite} to what 
we have found, implying that our results are not caused by any wavelength dependence in the
dust lane classification.

% why no SDSS-z 
\subsection{Signal-to-Noise Ratio}

Another possible observational bias that could play a role are differences in the signal-to-noise in the images of
the various SDSS filters. The SDSS{\em -z} images have notably lower S/N than the 
other filters due to the combined lower transmission of both filter and atmosphere. 
Therefore, the observed dust lane fraction is  low in the SDSS-{\em z} filter due to a combination of 
poor signal-to-noise and low contrast of a dust lane in the reddest filter. We note that the fraction for 
SDSS-{\em g, r} and {\em i} does not change as dramatically as from SDSS-{\em z} to SDSS-{\em i}. 
Combined with the poor consistency of the SDSS-{\em z} with SDSS-{\em g} (see \S \ref{s:ident} and 
Table \ref{t:sdss}), we discard SDSS-{\em z} as a reference. For the comparison with higher redshift, 
we show the dust lane fraction in SDSS-{\em g, r} and {\em i} filters as the redshift zero reference in Figure \ref{f:frac}. 

\subsection{Spatial Resolution}

The change in rest wavelength with redshift may not be the only effect biasing the dust lane fractions. 
At high redshift, galaxies are less resolved, which should make dust lanes more difficult to identify. 
This trend would counteract the bias due to bluer rest wavelengths. A minor effect could be that at lower redshifts, 
the slight smoothing of the image might aid in the identification of a galaxy-wide phenomenon such as a dust lane.

%FIGURE 5
\begin{figure}[h]
\begin{center}
\includegraphics[width=0.5\textwidth]{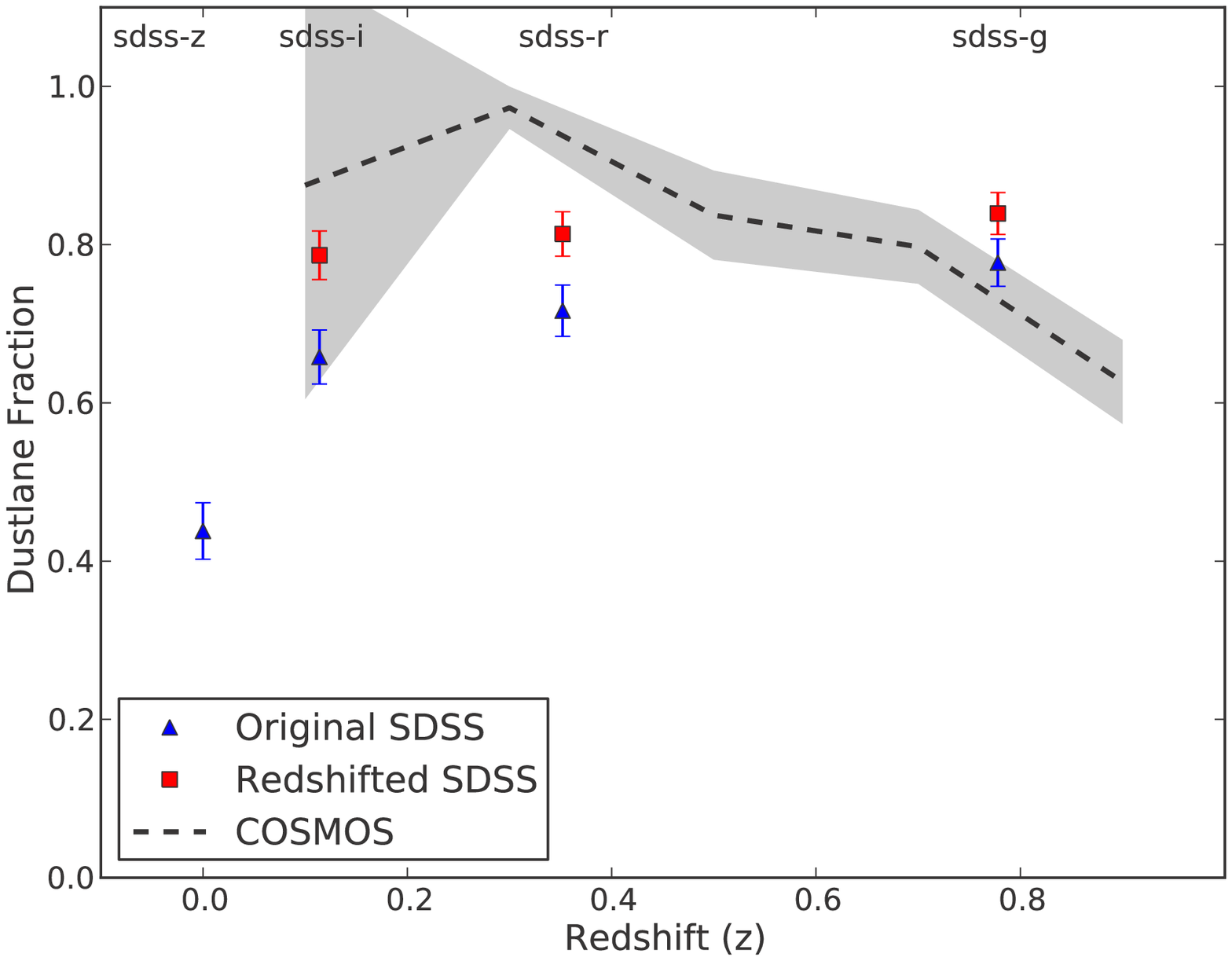}
\caption{The fraction of galaxies with identified dust lanes in the original SDSS images (blue triangle) and redshifted (dimmed, rebinned etc.) SDSS images as a function of redshift. The indicated redshift corresponds to the the restframe wavelength observed by the COSMOS {\em F814W} filter. The COSMOS fractions from Figure \ref{f:frac} are shown as a dashed line with uncertainties as the grey shaded area.}
\label{f:cal}
\end{center}
\end{figure}

% z-SDSS
To asses the total impact of these potential biases, in Figure \ref{f:cal}, we also plot the 
apparent dust lane fraction for the artificially redshifted SDSS-z images (squares). 
In addition to possible effects caused by changes in restframe wavelength, this test should also
account for the change in resolution and surface brightness.
Because there is little or no significant change in the dust lane fraction between $z\sim 0.1$ to $z \sim 0.8$,
we conclude that the reduced resolution and signal-to-noise at high redshift at most 
balances the increased sensitivity to dust lanes at shorter wavelengths. 

%FIGURE 6
\begin{figure}[h]
\begin{center}
\includegraphics[width=0.5\textwidth]{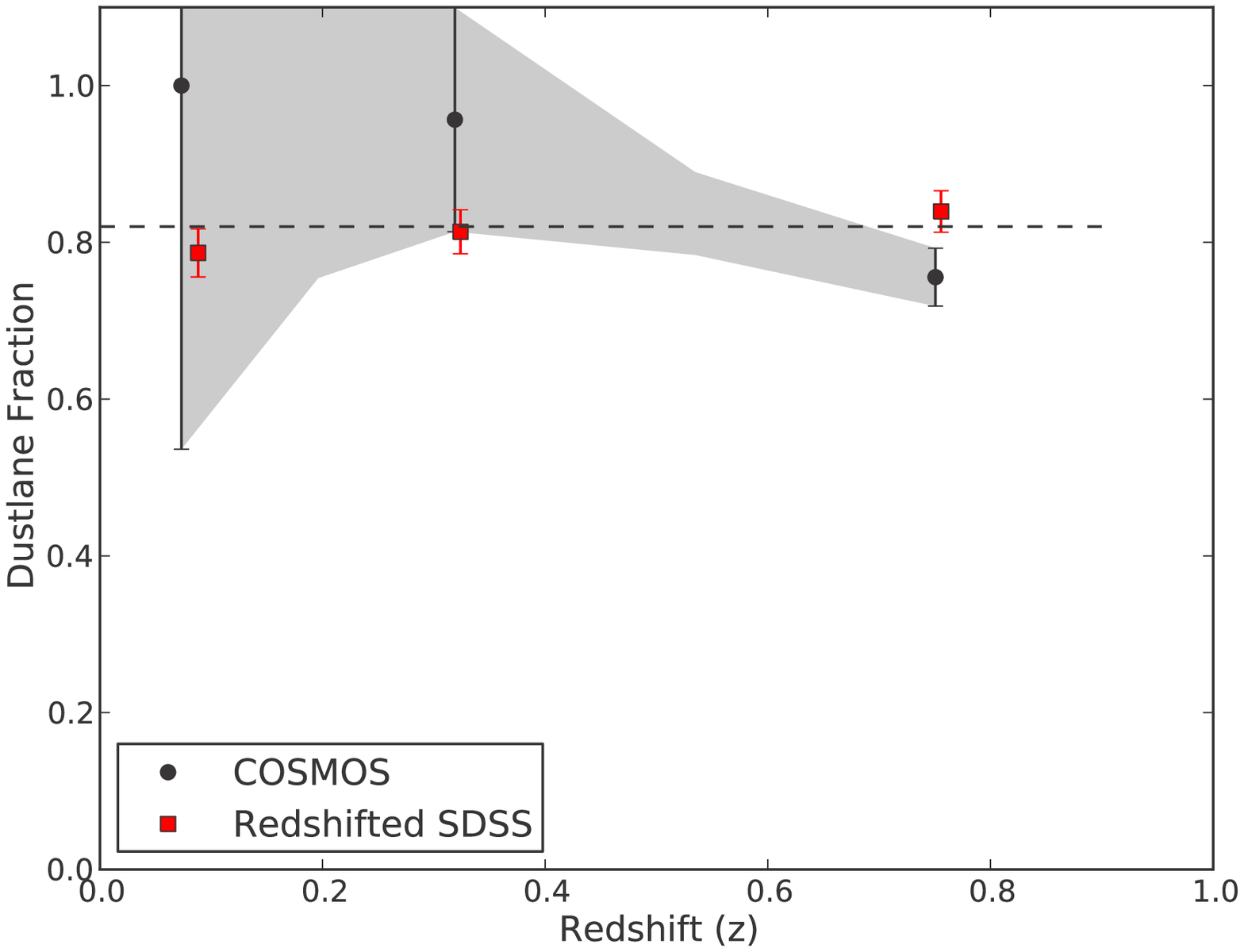}
\caption{The fraction of dust lanes in the redshifted SDSS and COSMOS, selecting only those COSMOS galaxies whose photo-z would place them in the bandwidth of the SDSS filter redshifted to the central wavelength of the {\em F814W} filter. Only the redshifted SDSS-{\em g, r} and {\em i} are considered. The respective widths of the shifted filters are marked with the horizontal errorbars. SDSS points are shifted by 0.005 redshift for clarity.} 
\label{f:check}
\end{center}
\end{figure}

\subsection{Filter Bandwidth}

As a final check for the reliability of the observed evolution in the dust lane fraction, we compare the dust lane fraction in the redshifted 
SDSS images ({\em g, r} and {\em i}) to that in the COSMOS galaxies whose redshifts fall within the SDSS filter's bandwidth. 
The {\em F814W} filter is much wider than the SDSS filters when these are shifted to the redshifts at which they correspond to the restframe wavelength.

From the various options for defining the bandwidth, we chose to select COSMOS galaxies whose photometric redshifts (photo-$z$) 
falls inside the bandwidth of the SDSS filter (after it was redshifted from its central wavelength to the central wavelength of the {\em F814W} filter.
Figure \ref{f:check} shows the results after resampling the COSMOS classifications in this way; black symbols for those intervals for which redshifted SDSS information is available. The gray area is all the COSMOS sample, including the intervening intervals. Table \ref{t:cosmos} lists the fraction of the COSMOS sample in the bins corresponding to the redshifted SDSS filters and dust lane fraction in the intervening intervals.

A constant fraction of 80\% dust lanes is consistent with almost the entire redshift range probed by COSMOS. Only close to $z\sim 0.75$ does the COSMOS fraction deviate by more than a standard deviation. The constant fraction is consistent with the redshifted SDSS images (within 95\% of each point). Thus, any evolution in the dust lane fraction is with only with 95\% ($1\sigma$) confidence but the slight rise with redshift appears unlikely due to the previously discussed observational biases.

One remaining observational biases is that the different redshifts in COSMOS predominantly probe different environments, i.e., field versus group or cluster. If the dust lane fraction depends on environment but not secular evolution, this could mimick the redshift evolution we see in Figure \ref{f:check}.

\begin{table*}
\caption{The fraction of COSMOS and redshifted SDSS galaxies in redshift bins corresponding to the SDSS filter widths: (1) redshift range, (2) corresponding rest-frame SDSS filter, (3) fraction of galaxies with dust lanes in redshifted SDSS sample, (4) fraction of dust lanes in COSMOS galaxies, (5) }
\begin{center}
\begin{tabular}{l l l l l l l }
Redshift          & 0.063 - 0.089      &  0.089 - 0.28       &  0.28 - 0.38        &  0.38 - 0.62        &  0.62 - 0.95        & \\ 
\hline
 SDSS              & sdss-i             &  \dots              &  sdss-r             &  \dots              &  sdss-g             & \\ 
 zSDSS frac        & 0.84 $\pm$ 0.03  &  \dots              &  0.81 $\pm$ 0.03  &  \dots              &  0.79 $\pm$ 0.03  & \\ 
\hline
 COSMOS frac       & 1.00 $\pm$ 0.46  &  0.94 $\pm$ 0.18  &  0.96 $\pm$ 0.14  &  0.84 $\pm$ 0.05  &  0.76 $\pm$ 0.04  & \\ 
\hline
 Tphot  	   &                    &                     &                     &                     &                     & \\ 
 \ \ $ < 4$            & 1.00 $\pm$ 0.46  &  1.00 $\pm$ 0.13  &  0.96 $\pm$ 0.14  &  0.88 $\pm$ 0.05  &  0.78 $\pm$ 0.04  & \\ 
 \ \ $ \geq 4$           & \dots  &  0.50 $\pm$ 0.44  &  \dots  &  0.67 $\pm$ 0.30  &  0.62 $\pm$ 0.23  & \\ 
 E(B-V)            &                    &                     &                     &                     &                     & \\ 
 \ \ $ < 0.1$            & \dots &  \dots  &  1.00 $\pm$ 0.46  &  0.50 $\pm$ 0.44  &  0.27 $\pm$ 0.29  & \\ 
 \ \ $ \geq 0.1$           & 1.00 $\pm$ 0.46  &  0.94 $\pm$ 0.18  &  0.95 $\pm$ 0.15  &  0.85 $\pm$ 0.05  &  0.80 $\pm$ 0.04  & \\ 
 log(M/M*)         &                    &                     &                     &                     &                     & \\ 
 \ \ $ < 0.1$            & 1.00 $\pm$ 0.46  &  1.00 $\pm$ 0.13  &  0.95 $\pm$ 0.16  &  0.82 $\pm$ 0.23  &  0.85 $\pm$ 0.05  & \\ 
 \ \ $ \geq 0.1$           & \dots  &  0.00 $\pm$ 0.45  &  1.00 $\pm$ 0.33  &  0.84 $\pm$ 0.06  &  0.70 $\pm$ 0.05  & \\ 
\hline
\end{tabular}
\end{center}
\label{t:cosmos}
\end{table*}%

\section{Analysis}
\label{s:analysis}

The COSMOS photometric redshift catalog of \cite{Mobasher07} uses fits to the Spectral Energy Distribution (SED) 
to provide several useful parameters for the COSMOS galaxy sample, in particular the photometric type ($T_{phot}$), 
internal reddening ($E[B-V]$), and stellar mass ($M^*$). 
Here, we explore possible dependencies of the dust lane fraction on these parameters. We apply the same bins as Figure \ref{f:check} and Table \ref{t:cosmos}.

% FIGURE 7
\begin{figure}[h]
\begin{center}
\includegraphics[width=0.5\textwidth]{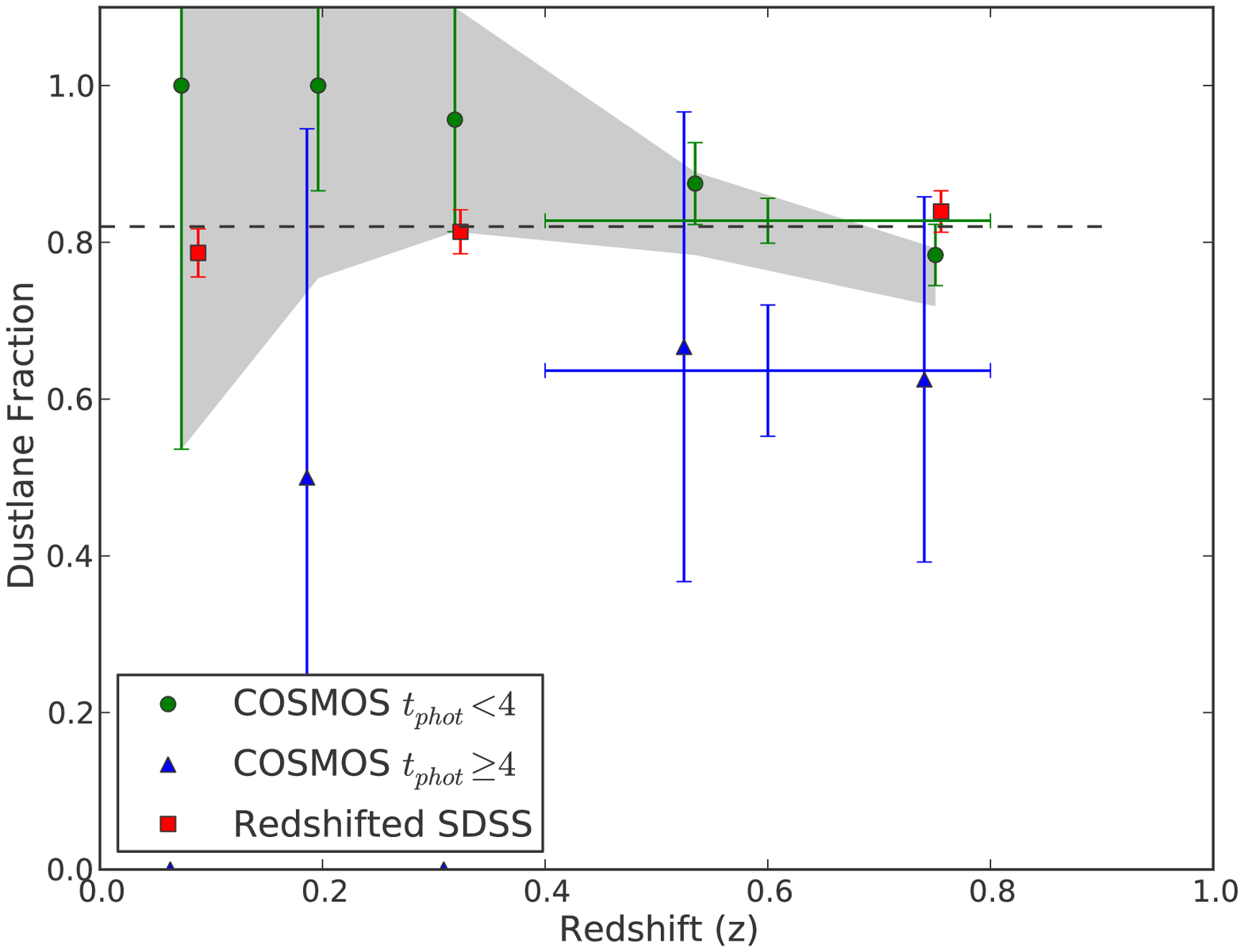}
\caption{The dust lane fraction of edge-on $L^*_V$ galaxies in the COSMOS survey, sorted by
star formation type ($T_{phot}$) from \cite{Mobasher07}. 
Plotted are the full COSMOS sample (gray area), early-type SEDs (\tphot $<4$, green circles) and late-type SEDs (\tphot $\ge4$, blue triangles), which are extinction-free starbursts (shifted by z=0.01 for clarity). The numbers are listed in Table \ref{t:cosmos}. The average for the last three redshift bins of COSMOS are indicated by the green and blue crosses without a symbol for early and late-type SEDs respectively.} 
\label{f:sf}
\end{center}
\end{figure}

% Star-Formation
The $T_{phot}$ parameter is a photometric Hubble type, separating red, early-type galaxies ($T_{phot}$ = 1) from the bluest, 
late-type galaxies ($T_{phot}$ = 6). The first four $T_{phot}$ types are \cite{Coleman80} templates, and the last two 
are from \cite{Kinney96}, these are starbursts with little internal extinction. In Figure \ref{f:sf}, we compare the dust lane fraction 
of quiescent ($T_{phot}<4$) to that of star-forming ($T_{phot} \ge4$) galaxies. The latter have sufficiently accurate statistics
only in the three highest redshift bins. 

We note that the SED fit allows only for a small amount of reddening of the stellar template \citep[$E(B-V) < 0.25$,][]{Mobasher07}, 
and in the case of edge-on spirals, one expect much higher levels of extinction. However, for a distant spiral with a clear dust lane, 
the SED fit is essentially made to the mostly undimmed extra-planar stellar population and the mid-plane population is too dim to be accounted for. Therefore, some of the quiescent SED galaxies may, in fact, be hosting some star-forming regions deeply embedded 
in the dusty structures.

The more quiescent SED galaxies ($T_{phot}< 4$), which dominate the COSMOS population, show the gradual increase in dust lane fraction that is nevertheless consistent with a constant fraction of 80\%, similar to the full COSMOS sample and redshifted SDSS images. 
The star-forming galaxies ($T_{phot}=4-6$), on the other hand, have a consistently lower 
dust lane fraction, although the constant fraction of 80\% is still within one standard deviation due to the much lower statistics for star-bursting galaxies. By combining the last three redshift bins in the COSMOS sample, the fraction of dustlanes in starbursting galaxies is significantly lower than the quiescent galaxies.
A lower dust lane fraction is consistent with the notion that greatly enhanced specific star-formation (a starburst) destroys or inhibits the formation of a dust lane.
However, a lower level of star-formation can co-exist with a dust lane as the star-formation would inherently be deeply embedded.
To give a definite answer, one would need a star-formation rate estimate based on the full SED (including sub-mm to estimate the embedded star-formation) and much 
improved statistics. To summarize, Figure \ref{f:sf} is consistent with a constant dust lane fraction for both populations of galaxies.

% FIGURE 8
\begin{figure}[h]
\begin{center}
\includegraphics[width=0.5\textwidth]{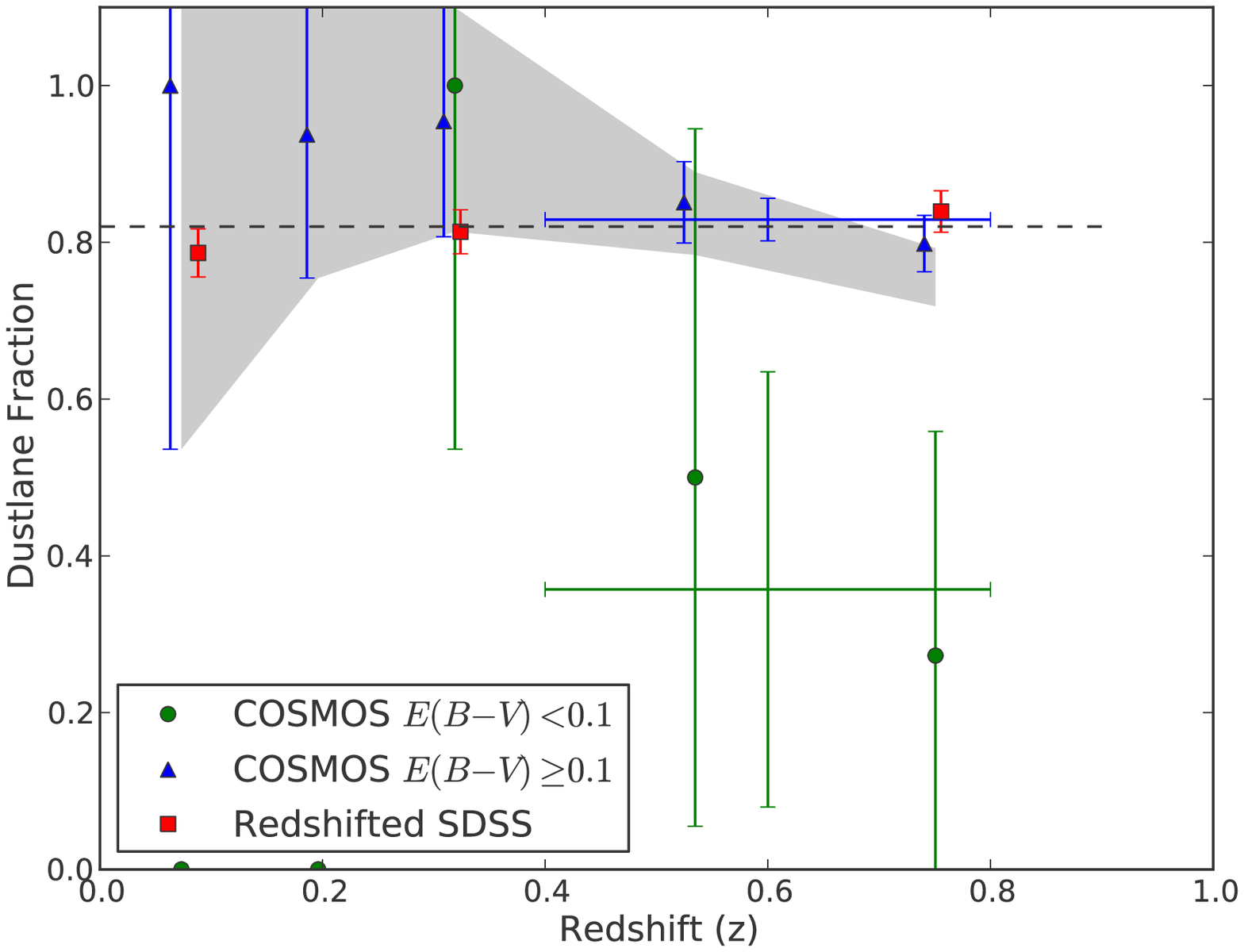}
\caption{The dust lane fraction of edge-on $L^*_V$ galaxies in the COSMOS survey, sorted by internal dust reddening of 
the SED as derived by \cite{Mobasher07}. Plotted are the combined COSMOS sample (gray area), 
reddened galaxies ($E[B-V]>0.1$, blue triangles), and galaxies with little or no reddening ($E[B-V] < 0.1$, green circles, shifted by z=0.01 for clarity). The average for the last three redshift bins of COSMOS are indicated by the green and blue crosses without a symbol for low and high reddening respectively.
At low redshift, effectively all galaxies display evidence of reddening in their SED, but there is a small 
population of galaxies at higher redshift that does not show any evidence for reddening or a dust lane. 
The numbers are listed in Table \ref{t:cosmos}.   } 
\label{f:ext}
\end{center}
\end{figure}

% E(B-V)
The photometric fits of \cite{Mobasher07} also report the internal dust reddening of the SED template. Figure \ref{f:ext} 
shows the dust lane fraction for reddened ($E[B-V] > 0.1$) and little or not-reddened galaxies  ($E[B-V]  \leq 0.1$). 
The majority of galaxies in the COSMOS sample show reddening of their SED. The small sub-sample of galaxies 
that do not show reddening also has a lower incidence of dust lanes, especially at z=0.7. To counter the poor statistics, we combine the last three redshift bins in COSMOS (crosses in Figure \ref{f:ext}). The low- or no-extinction population is offset from the general COSMOS population with only 30\% showing a dust lane.

This small subsample with no dust reddening is surprising because edge-on spirals are the objects in which one would expect dust reddening to be most pronounced, with the possible exception of (U)LIRGS or merging systems. 
Yet, there is a small population of edge-on disks at higher redshift with little reddening and no apparent dust lane (Table \ref{t:cosmos}).
If the dust lane is dispersed in these disks and there are dusty structures throughout these galaxy disks, 
(similar to the lower mass disks in D04), one would expect some reddening of the stellar template.
We speculate that in these cases, the galaxy's dusty, molecular disk is much smaller than the stellar one, possibly after 
losing much of their ISM to the surrounding IGM through, for instance, a galactic fountain \citep[e.g.,][]{Sturm11} or stripping.

% FIGURE 9
\begin{figure}[h]
\begin{center}
\includegraphics[width=0.5\textwidth]{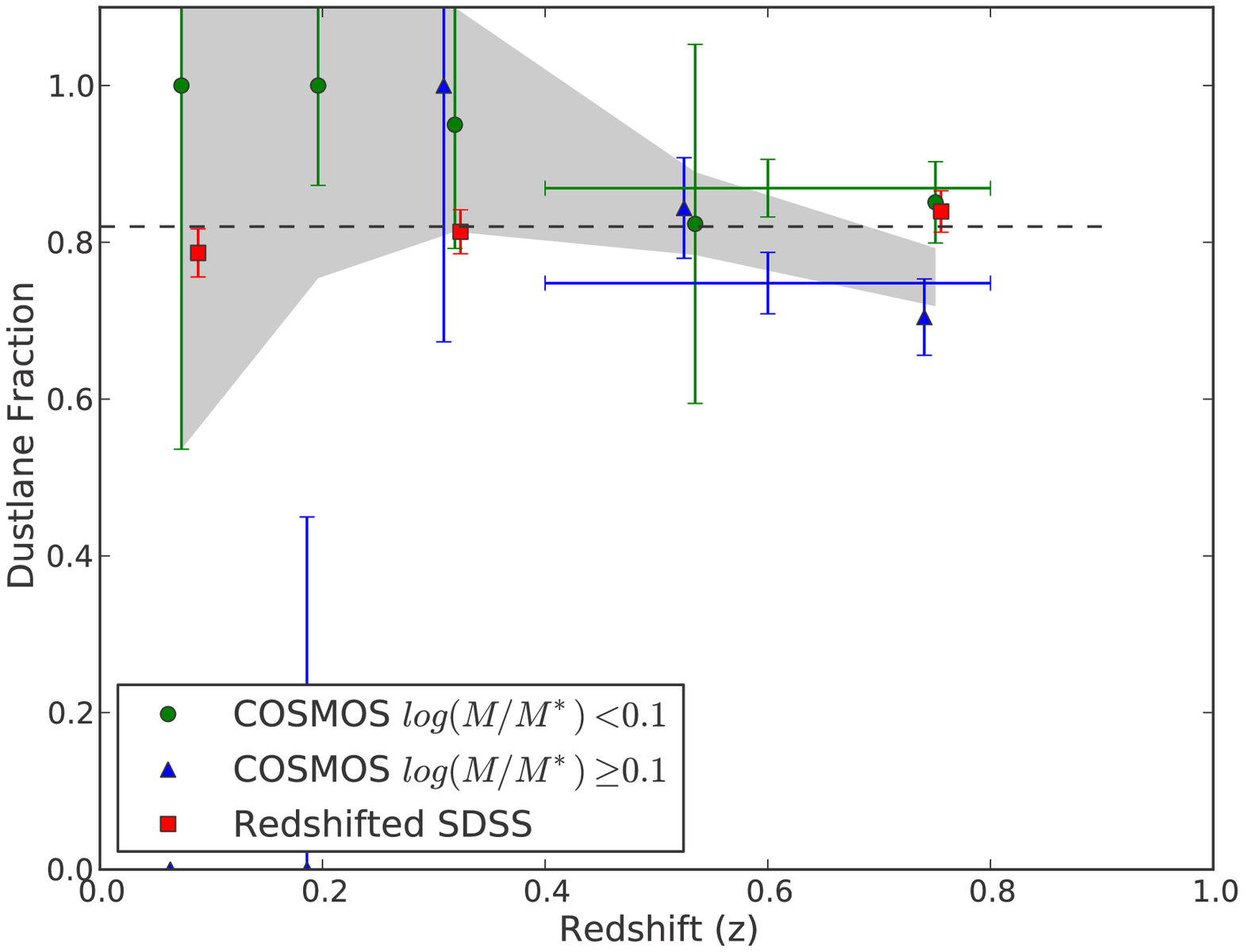}
\caption{The dust lane fraction of edge-on $L^*_V$ galaxies in the COSMOS survey, sorted by stellar mass 
as derived by \cite{Mobasher07}. Lower and higher stellar masses are the green circles and blue triangles, respectively, the latter shifted by z=0.01 for clarity. The average for the last three redshift bins of COSMOS are indicated by the green cross for low-mass and blue cross for high-mass galaxies.
Both mass ranges display similar fractions of dust lanes out to $z\leq0.6$, with a deviation at z=0.7 (Table \ref{t:cosmos}). } 
\label{f:mass}
\end{center}
\end{figure}

% Mass
Finally, the dust lane fractions at different stellar masses are shown in Figure \ref{f:mass}. The high and low mass samples are both consistent with a constant fraction of dust lanes (80\%) with the exception of the points at z=0.7. Here the {higher} mass galaxies display a lower dust lane fraction, while the lower mass galaxies are consistent with the SDSS and the constant fraction of 80\%. If we average over the last three redshift bins to improve statistics (crosses with no symbols in Figure \ref{f:mass}), the discrepancy between high and low-mass galaxies remains.
This implies a reversal of the relation between mass and occurrence of dust lanes seen in D04. However, both samples consist of massive galaxies as per the definition of D04 ($v_{rot} > 120$ km/s). One could speculate that the a fraction of the massive spirals in COSMOS at z=0.7 have recently undergone a major merger (as most massive spirals are predicted to have since $z\sim1$). Thus, in some massive disks, a few of the dust lanes are now missing. The break in dust lane fraction observed by D04 occurs below a stellar mass of 0.1 $M^*$. 
The starburst galaxies in Figure \ref{f:sf} could be those massive disks that are missing their dust lane.

\section{Discussion}
\label{s:disc}

The main assumption in D04 and this paper is that dust traces the coldest, often molecular, phase of the ISM. This assumption is 
based on both observational data \citep[e.g.,][Holwerda et al. {\em submitted}]{Allen86,Vogel88, Rand95, Planck-Collaboration11a}, and theoretical arguments, which suggest that dust and molecular gas have slow relative drift velocities \citep{Weingartner01b}. Furthermore, dust acts as a catalyst for molecular hydrogen ($H_2$) production \citep{Cazaux04b}, and dust shields the molecular gas from dissociation by ultra-violet radiation. 
Our finding that the dust lane fractions seems independent of redshift out to z=0.7 (Figure \ref{f:check}) therefore implies that 
the global state of the ISM in large spiral disks has not changed over the past 7 Gyr. We emphasize that this statement is valid only for the galaxy population 
as a whole, and that {\em in individual disks}, dust lanes may well be destroyed and re-formed on short time-scales. 
Taken at face value, however, the constant fraction over time suggests that dust lanes are a long-lived phenomena.
% However, the gradual increase suggest that dust lanes are (re)coalescing after the epoch of high specific star-formation at $z\sim1$.

The observation that approximately the same number of massive galaxies have thin dusty molecular disks at z=0.8 has significant implications for models of their emission and the supernovae they host. 
For instance, SED models of local massive edge-on systems are still suitable for explaining the characteristics of these higher redshift systems \citep[e.g.,][Holwerda et al. {\em submitted}]{Xilouris99,dirty1,dirty2,Bianchi07,Bianchi08, Baes03, Baes10a, Bianchi11, Holwerda11iau, Schechtman-Rook12a, Schechtman-Rook12b}.
The SNIa lightcurve fits of Dark Energy Cosmology depend on a host galaxy dust extinction prior based on a flat dusty spiral disk \citep[e.g.,][]{Hatano98,Lah07,Holwerda08a}. For those SNIa found in massive spiral disks (likely the majority), the locally derived prior would be applicable out to $z\sim0.8$.

New {\em Herschel} results suggest a strong evolution in the population of dusty galaxies \citep[e.g.,][]{Dye10, Clements10}. 
There are many more sub-mm luminous galaxies at higher redshift then there are today, a result of both enhanced dust production and higher irradiation due to star-formation. 
The {\em Herschel} results show how the dusty ISM's {\em illumination} rapidly evolves with time.
This paper, in contrast, focuses on the evolution of dusty ISM's {\em morphology} in massive spiral disks and finds little or no evidence for evolution since z=0.7. 
Both dust illumination and geometry are key ingredients for models explaining the population of luminous infra-red and sub-mm galaxies observed at high redshift.

Because we select galaxies that are expected to passively evolve into $M^*$ galaxies \citep[see][]{Capak04,Sheth08a}, these systems are the massive haloes and disks  expected to have collapse first. They have produced a large fraction of their stellar mass at the epoch of observation \citep{Bell03c,Ilbert10}, and the mass surface density of all components combined (including dark matter, molecular and atomic gas) is in a similar vertical balance with turbulence as it is today i.e., sufficient to collapse the ISM into a thin dust lane. Since the specific star-formation is a factor three higher for the same stellar mass \citep{Noeske07a,Noeske07b}, it implies that the total ISM surface density must be higher to compensate. The proportions of the ISM components depend on the local pressure \citep[e.g.,][]{Obreschkow09e, Obreschkow09b} and we speculate that increase in ISM surface density and star-formation feedback both point to a more important molecular component at this epoch.

% and together with the other disk components (dark matter, molecular and atomic hydrogen) contribute enough to the disk surface density to collapse it vertically into a thin disk in most cases. In what proportion these other components contribute depends on the pressure balance between them \citep[see][]{Obreschkow09e, Obreschkow09b}.

% implication for the maximum disk at higher redshift.
%The high disk mass surface density implied by the existence of a dust lane and the low stellar masses inferred from SED fits indicate that the maximum disk conjecture (i.e, that the observable baryons --mostly stars-- fully explain the rotation curve in the stellar part of the disk) may need some adjustment for \lv \ galaxies in earlier epochs. The necessary baryons may well be in the interstellar gas, not in the stars. Alternatively, the dark matter may still play a more significant role in the inner disk at those epochs.

\section{Conclusions}
\label{s:concl}

By identifying dust lanes in highly inclined spiral disks of \lv \ galaxies in the SDSS and COSMOS surveys, we conclude the following:

\begin{enumerate}
\item Our estimate (80\%) for the dust lane fraction in local galaxies, derived from SDSS-{\em g} images, is consistent with previous results from \cite{Dalcanton04} (Figure \ref{f:frac}).

\item Any evolution in the dust lane fraction with redshift derived from COSMOS images can not be caused by the effects of varying rest-frame wavelength or spatial resolution 
for galaxies at higher redshift (Figure \ref{f:cal}). Taken in the same filter bandwidth, the redshifted SDSS and the COSMOS classifications are both consistent with a {\em constant} 
dust lane fraction of about 80\% of galaxies since $z\sim 0.8$ (Figure \ref{f:check}).

\item The majority of \lv \ galaxies in COSMOS (75\% or more) display a dust lane over the last 7 Gyr. The observed fractions in COSMOS are consistent with a 
constant fraction of 80\%, i.e., with no evolution since z=0.7. At z=0.8, close to the redshift limit of our survey, the dust lane fraction appears slightly
lower, hinting at some evolution beyond z=0.7 (Figure \ref{f:check}). 

 \item Star-forming galaxies (SED type \tphot \ = 4-6) seem to have a lower dust lane fraction ($f\sim$50\%) than more quiescent SED types (\tphot$<4$, $f\sim 80$\%), 
 consistent with the picture that a high specific star-formation rate injects enough energy into the ISM disk that the additional turbulence prevents the vertical collapse 
 of the dusty molecular ISM into a dust lane (Figure \ref{f:sf}). However, we caution that due to the small number of star-forming galaxies, the associated error bars are
rather large, so this result is somewhat tentative.

\item A small population of edge-on spirals with no reddening or dust lane is evident at higher redshift (Table \ref{t:cosmos}, Figure \ref{f:ext}). We speculate that their dust disks do not extend far enough into the spiral disk to project a dust lane.

\item At the highest redshift ($z\sim0.8$), the lower mass galaxies ($log(M/M^*) < 0.1$) show a slightly higher fraction of dust lanes than the high mass galaxies ($log(M/M^*) > 0.1$, Figure \ref{f:mass}). This may be due to evolution, a change in environment or a higher contribution of starbursting galaxies at higher redshift to the massive galaxy population.

\item Many of the $L^*_V$ disks have already accumulated enough surface density 7 Gyr ago in all components combined  to vertically collapse the cold, dusty ISM into a thin lane. This implies SED or SNIa host galaxy models using this geometry based on local results are applicable to higher redshift ($z< 0.8$)

% Because there is no difference in dust lane fraction for large and small stellar masses (Figure \ref{f:mass}), the mass of this thin disk must be in the other components; e.g., dark matter, molecular and atomic hydrogen.
\end{enumerate}

% CANDELS
Future work on the fraction of dust lanes in edge-on spiral disks will be possible with the CANDELS survey which will provide three color images of five fields. The near-infrared imaging with HST allows to probe further along the mass range, possibly down to the mass where dust lanes disappear in these distant galaxies.
In the Sloan Digital Survey, we hope to use the classification by the GalaxyZoo2 volunteers to improve on the results of D04, \cite{Obric06} and this paper for the local Universe.
Lastly, edge-on galaxies can be observed with the Atacama Large Millimetre Array ({\em ALMA}) to probe both the dynamics and the width of the molecular phase in spirals.

\section*{Acknowledgements}

The authors would like to thank the anonymous referee, whom's comments led to a substantially improved manuscript, and K. Sheth for the useful discussions in framing the samples. We acknowledge support from HST AR grant GO-1274.
Funding for the SDSS and SDSS-II has been provided by the Alfred P. Sloan Foundation, the Participating Institutions, the National Science Foundation, the U.S. Department of Energy, the National Aeronautics and Space Administration, the Japanese Monbukagakusho, the Max Planck Society, and the Higher Education Funding Council for England. The SDSS Web Site is http://www.sdss.org/.
The SDSS is managed by the Astrophysical Research Consortium for the Participating Institutions. The Participating Institutions are the American Museum of Natural History, Astrophysical Institute Potsdam, University of Basel, University of Cambridge, Case Western Reserve University, University of Chicago, Drexel University, Fermilab, the Institute for Advanced Study, the Japan Participation Group, Johns Hopkins University, the Joint Institute for Nuclear Astrophysics, the Kavli Institute for Particle Astrophysics and Cosmology, the Korean Scientist Group, the Chinese Academy of Sciences (LAMOST), Los Alamos National Laboratory, the Max-Planck-Institute for Astronomy (MPIA), the Max-Planck-Institute for Astrophysics (MPA), New Mexico State University, Ohio State University, University of Pittsburgh, University of Portsmouth, Princeton University, the United States Naval Observatory, and the University of Washington.
This research has made use of the NASA/ IPAC Infrared Science Archive, which is operated by the Jet Propulsion Laboratory, California Institute of Technology, under contract with the National Aeronautics and Space Administration.
This research has made use of the NASA/IPAC Extragalactic Database (NED) which is operated by the Jet Propulsion Laboratory, California Institute of Technology, under contract with the National Aeronautics and Space Administration. 
This research has made use of NASA's Astrophysics Data System.

%\bibliographystyle{apj} 
%\bibliography{/Users/bholwerd/Desktop/Science/Bib/Bibliography}

\end{document}